\documentclass[12pt,preprint]{aastex}
\usepackage{times}

\shorttitle{Subsurface Properties of Active Regions}
\shortauthors{Zhao et al.}

\begin{document}
\title{High Resolution Helioseismic Imaging of Subsurface Structures and 
Flows of A Solar Active Region Observed by {\it \bf Hinode}}

\author{Junwei Zhao, Alexander G. Kosovichev}
\affil{W.~W.~Hansen Experimental Physics Laboratory, Stanford University,
Stanford, CA94305-4085}
\and
\author{Takashi Sekii}
\affil{National Astronomical Observatory of Japan, 2-21-1 Osawa, 
Mitaka, Tokyo 181-8588, Japan} 

\begin{abstract}
We analyze a solar active region observed by the {\it Hinode} \ion{Ca}{2}~H
line using the time-distance helioseismology technique, and infer 
wave-speed perturbation structures and flow fields beneath the active 
region with a high spatial resolution. The general subsurface wave-speed 
structure is similar to the previous results obtained from {\it SOHO}/MDI 
observations. The general subsurface flow structure is also similar, and 
the downward flows beneath the sunspot and the mass circulations around 
the sunspot are clearly resolved. Below the 
sunspot, some organized divergent flow cells are observed, and these structures 
may indicate the existence of mesoscale convective motions. Near the light 
bridge inside the sunspot, hotter plasma is found beneath, and flows 
divergent from this area are observed. The {\it Hinode} data also allow us 
to investigate potential uncertainties caused by the use of phase-speed
filter for short travel distances. Comparing the measurements with and 
without the phase-speed filtering, we find out that inside the sunspot, 
mean acoustic travel times are in basic agreement, but the values are 
underestimated by a factor of $20-40\%$ inside the sunspot umbra
for measurements with the filtering. The initial acoustic tomography
results from {\it Hinode} show a great potential of using high-resolution
observations for probing the internal structure and dynamics of sunspots.

\end{abstract}

\keywords{Sun: helioseismology --- Sun: sunspots --- Sun: interior}

\section{Introduction}
Deriving subsurface structures and flow fields of solar active regions is 
one of the major topics for local helioseismology studies. Using \ion{Ca}{2}~K 
observations made at the geographic South Pole in early 1990s, \citet{duv96} 
found the first evidence of downdrafts below active regions through 
measuring acoustic travel times by employing the time-distance helioseismology 
technique. Later inversions \citep{kos96} using these travel time 
measurements found strong converging downflows and areas of the excess of 
sound-speed beneath growing active regions. With the availability 
of {\it Solar and Heliospheric Observatory} / Michelson Doppler Imager
\citep[{\it SOHO}/MDI;][]{sch95} Doppler observations that are seeing-free and 
more suitable for local helioseismology studies, more investigations 
of the sunspot's subsurface structures and flow fields have been carried
out by use of various theoretical models: ray-path approximation 
\citep[e.g.,][]{kos00, zha01}, Fresnel-zone approximation 
\citep[e.g.,][]{jen01, cou04}, and Born approximation \citep[e.g.,][]{cou06}. 
Despite the use of different approaches of calculating travel time
sensitivity kernels, the results remain largely the same, i.e., for 
active region subsurface structures, a negative sound-speed perturbation 
was found up to 5 Mm below the photosphere, and a positive perturbation 
was seen below that depth. For subsurface flow fields, converging downflows 
were inferred from the photosphere to about 5 Mm beneath it, and divergent 
flows were found below this layer. Another local helioseismology technique, 
ring-diagram analysis, although could not provide subsurface flow maps 
with such a high spatial resolution, gave results that were in general 
agreement \citep{hab04, kom05}. Some direct comparisons of these two
techniques were also performed \citep{hin04}.

However, on the other hand, these results were not uncontroversial. 
By measuring acoustic phase shifts in sunspot penumbra using MDI
Doppler observations when sunspots were located near the solar disk 
limb, \citet{sch05} demonstrated that the measured acoustic travel 
times varied around sunspot penumbra when the sunspot was located near
the limb, and believed this effect was caused by the inclined magnetic field.
While acknowledging the existence of such an effect in oscillation signals
observed in Dopplergrams, \citet{zha06} found out that such an effect
did not exist in the oscillations observed in MDI continuum intensity. 
Additionally, \citet{raj06} showed that the phase-speed filtering in the
time-distance acoustic travel time measurements would also introduce 
errors in active regions where oscillation amplitudes were suppressed.
Using numerical simulations \citet{par08} confirmed this effect, and 
concluded that this effect led to underestimation of the mean travel
times at short distances. Some recent attempts of using different 
filtering procedures, e.g., ridge filtering \citep{bra08}, showed that 
different filtering might give different measurements. Thus, it is 
certainly worthwhile examining how filters would change measurement results.
On the other hand, it is also arguable whether those time-distance 
observed acoustic travel time variations in solar magnetic areas 
could be interpreted as interior sound speed anomalies, or were caused 
by some surface magnetism effects such as showerglass effect \citep{lin05a, 
lin05b}. Furthermore, it is also not clear how the interaction of acoustic 
waves and magnetic field would effect the phase of these waves, hence the
travel time variations in the measured acoustic travel times. It
would be very difficult to interpret the measurements if acoustic
waves experience some phase shifts at the boundary of unmagnetized
and magnetized areas \citep{cal09}. 

High spatial resolution observation of sunspots by the Solar Optical 
Telescope \citep[SOT;][]{tsu08} onboard the Japanese solar spacecraft 
{\it Hinode} \citep{kos07} not only gives us another possibility of 
investigating subsurface properties of solar active regions in addition to 
the already existing helioseismology instruments, but also provides an 
unprecedented high spatial resolution that has enabled some studies that
were improbable using these existing instruments. In this paper, we 
analyze an active region observed by {\it Hinode} utilizing the time-distance 
helioseismology technique, and infer the wave-speed profiles and 
flow fields beneath this region, and present these results in \S3.
We also analyze the acoustic signals without applying any filters
except filtering out solar convection and $f$-modes,
and investigate how the phase-speed filtering effects our 
measurements. Such analyses can hardly be carried out by use of MDI 
high resolution observations, and these results are presented in \S4. In \S5,
we discuss and summarize our results.

\section{Observation and Data Analysis}
\begin{figure}
\epsscale{1.0}
\plotone{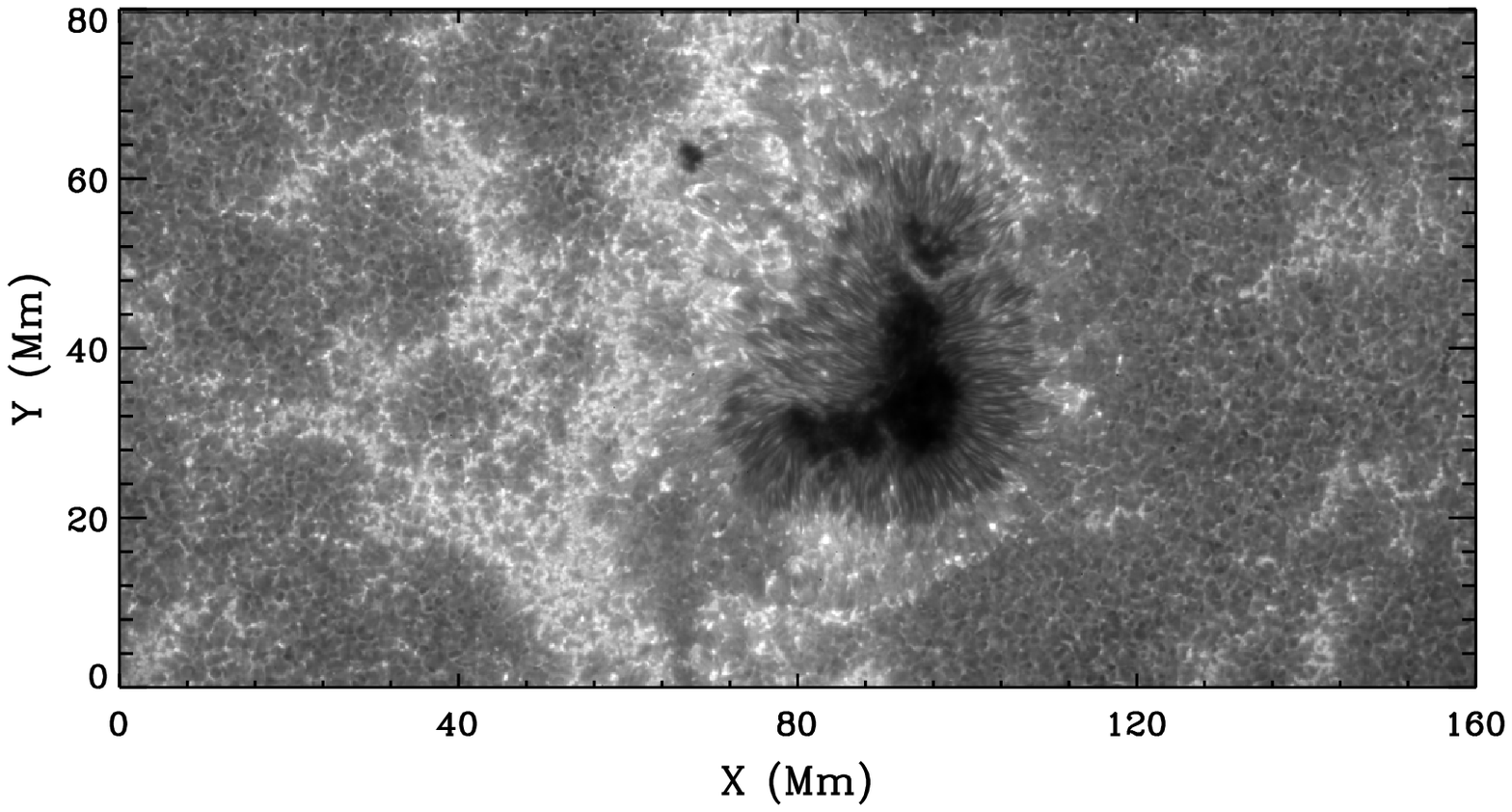}
\caption{A \ion{Ca}{2}~H image of the studied region from the 
{\it Hinode}/SOT. } 
\label{spot}
\end{figure}

The sunspot inside active region NOAA AR10953 was observed by 
{\it Hinode}/SOT \ion{Ca}{2}~H line continuously for a total of 
738 minutes, from 17:00UT May 2 through 05:58UT May 3, 2007, with a cadence 
of approximately 1 min. The cadence was not uniform, but had a variation 
of a fraction of 1 sec, and this was corrected by interpolating data into 
a uniform time grid of 1 min. Each image, $2048 \times 1024$ pixels in 
size, has a spatial resolution of $0.108\arcsec$/pixel, and a field of 
view of approximately $160 \times 80$ Mm. An image from this 
observation is shown in Figure~\ref{spot}. We perform a 
$2 \times 2$ rebin before time-distance analysis, and the resultant 
resolution is corresponding to 0.156~Mm/pixel. During this $12+$ hours 
observation, the pointing of SOT was not sufficiently stable, and this
is corrected by shifting images to keep the sunspot center in a fixed 
location during the whole sequence. 

As demonstrated by \citet{sek07} and \citet{nag09}, the time-distance 
helioseismology technique works well with the {\it Hinode} \ion{Ca}{2}~H 
data. We apply the similar measurement and inversion procedure on this active 
region. Seven different time-distance measurement annuli are used,
and the annulus ranges are $4.0 - 6.6$, $6.5 - 9.7$, $9.2 - 12.5$, 
$12.0 - 15.9$, $15.1 - 18.9$, $18.5 - 22.3$, and $21.6 - 25.4$ Mm, 
respectively. Because of the limited field of view,
longer annuli cannot be used in measuring this dataset, and this restricts 
us from reaching deeper interiors. When we fit the cross-covariance inside 
the sunspot umbra and penumbra some misfittings occur, but the number 
of misfittings is insignificant, and this should not alter our analysis 
and final results. The misfitted pixels are replaced by averaged values 
from their neighboring pixels.

After the acoustic travel times are measured, inversions are performed to 
infer the subsurface magnetoacoustic wave-speed structures and flow fields. 
The inversion technique, sensitivity kernels, and the final averaging 
kernels are essentially the same as or similar to those presented by 
\citet{zha07} except the slightly different annulus radii and depth coverage.

\section{Subsurface Structure and Flow Field}
\subsection{Subsurface Wave-Speed Structure}
\begin{figure}
\epsscale{1.0}
\plotone{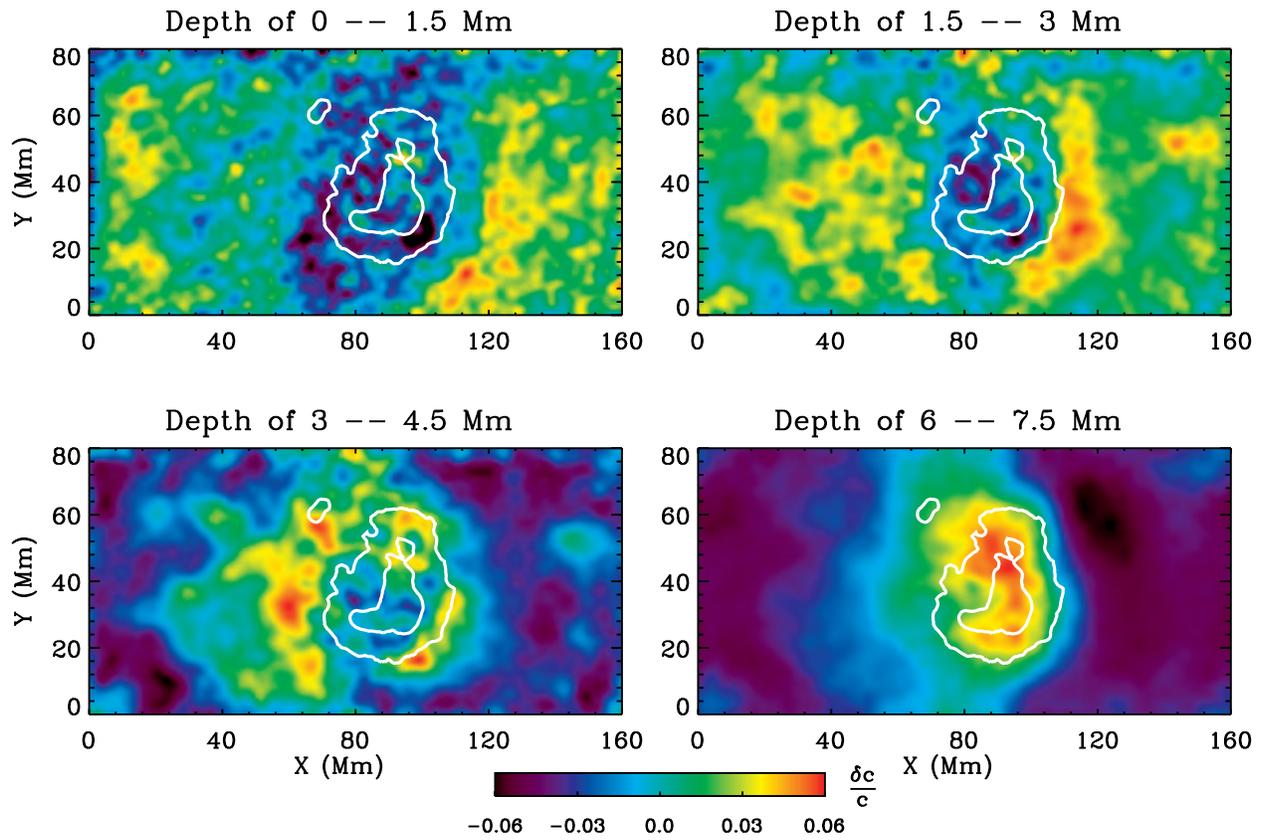}
\caption{Sound speed perturbation at different depths. The white contours in
each panel indicate the boundaries of sunspot umbra and penumbra. }
\label{cs_horiz}
\end{figure}

A few selected images of the inverted wave-speed perturbations 
at different depths are displayed in Figure~\ref{cs_horiz}. At 
depths shallower than $\sim3$ Mm, the wave-speed perturbation
is dominantly negative inside the sunspot area, although small areas 
of positive perturbations exist, in particular, close to a light bridge
in the upper umbra area. At the depth of $3.0 - 4.5$ Mm, negative and 
positive wave-speed perturbations are mixed. Below approximately 
6 Mm, the perturbations are predominantly positive inside the
sunspot area. Please note that all depths here are relative to where 
the \ion{Ca}{2}~H line forms and the acoustic signals are observed, i.e., 
approximately 250 km above the solar photosphere \citep{car07}.

\begin{figure}
\epsscale{0.8}
\plotone{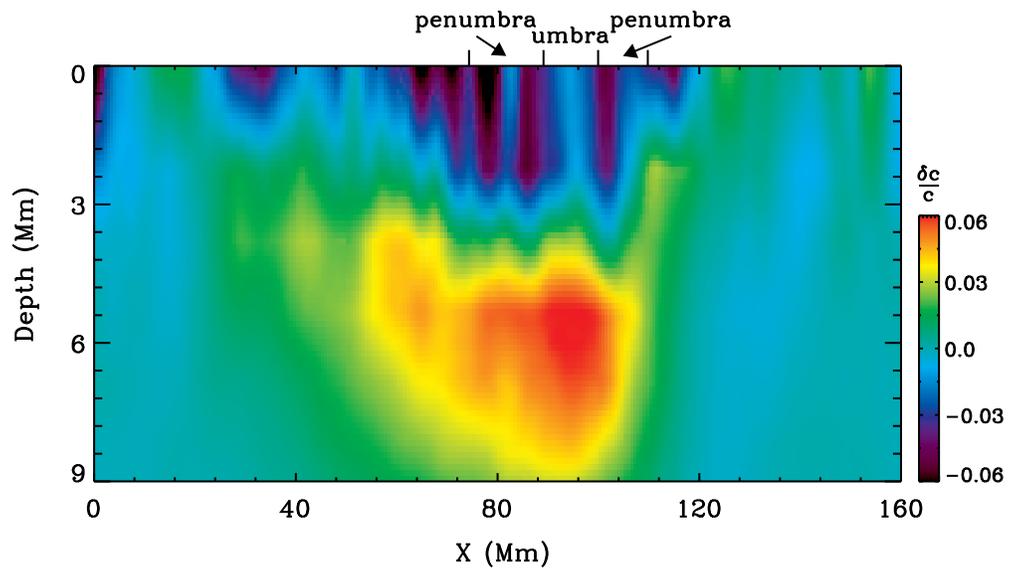}
\caption{A vertical view of sound speed perturbation beneath the active 
region. }
\label{vert_cs}
\end{figure}

\begin{figure}
\epsscale{0.8}
\plotone{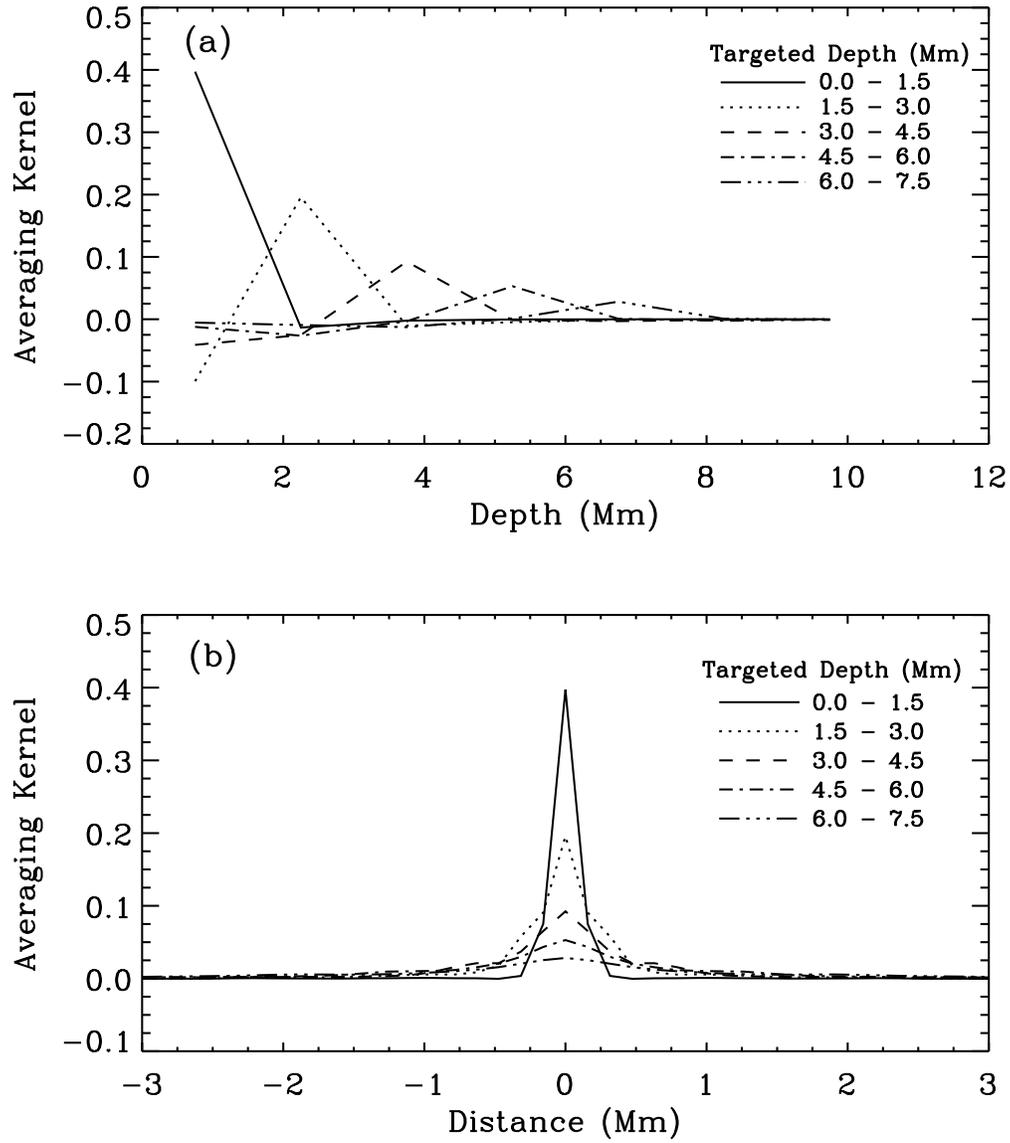}
\caption{Averaging kernels, as functions of (a) depth and (b) horizontal
distance, for different targeted depths obtained from acoustic wave 
speed perturbation inversion. }
\label{ave_kernel}
\end{figure}

A vertical cut through the center of the sunspot along the East-West
direction at $Y = 39$ Mm, as presented in Figure~\ref{vert_cs},
shows that the wave-speed anomalies extend about half of the sunspot 
size beyond the sunspot penumbra horizontally. In the vertical direction, 
the negative wave-speed perturbation extends to a depth of $\sim4$ Mm. 
The positive perturbation is about 9 Mm deep, but it is not clear whether 
it extends further, because our inversion cannot reach deeper layers 
due to the limited field of view. The averaging kernels associated
with this inversion are shown in Figure~\ref{ave_kernel}. 

\begin{figure}
\epsscale{0.9}
\plotone{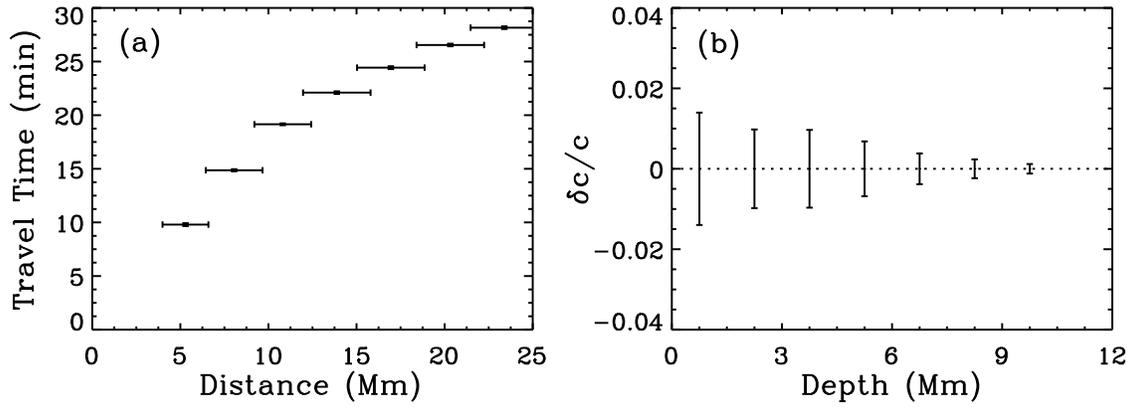}
\caption{(a) Measured mean travel times obtained in a piece of quiet Sun.
Horizontal error bars indicate the measurement annulus ranges, and vertical
error bars indicate standard deviations from this area. (b) Error 
estimates from acoustic wave speed perturbations at each targeted depth.}
\label{err_est}
\end{figure}

In the quiet Sun, the measured mean acoustic travel times often
appear to be relatively constant with little correlation with 
supergranulation structures. Hence we can estimate our travel time
measurement uncertainties by computing standard deviations in a 
region of the quiet Sun. The errors in the inverted wave speed 
structures can then be estimated by convolving the measurement uncertainties
with the averaging kernels. These error estimates are shown in 
Figure~\ref{err_est}. It is found that the errors decrease and the 
averaging kernels become broader with the increase of depth. This 
is a general property of helioseismic inversions, and the error 
magnitude and averaging kernel width depend on the choice of 
regularization parameters. In this case, the parameters
are chosen to provide a sufficient smoothing of the regularized
solution at different depths. This results in broader avergaing
kernels and smaller error estimates in deeper layers.

This picture of sunspot's subsurface wave-speed perturbation is made 
from intensity observations, rather than more commonly used Doppler 
velocity data. In spite of this, 
the subsurface wave-speed structure is remarkably similar to the previous
results obtained using MDI Dopplergrams and various types of the acoustic 
sensitivity kernels \citep[e.g.,][]{kos00, cou06, zhar07}. It is also 
noticeable that no obvious effect caused by the inclined magnetic field 
is visible in the inverted sound speed perturbation map, 
although the sunspot is located quite a distance away from the solar disk 
center. This is in agreement with the conclusion of that the solar 
oscillation signals observed in intensity are not affected by inclined 
magnetic fields \citep{zha06}. Unfortunately, a direct comparison with MDI is
not possible because of the lack of simultaneous uninterrupted observations
for this sunspots or other sunspots.

\subsection{Subsurface Flow Field}
\begin{figure}
\epsscale{0.59}
\plotone{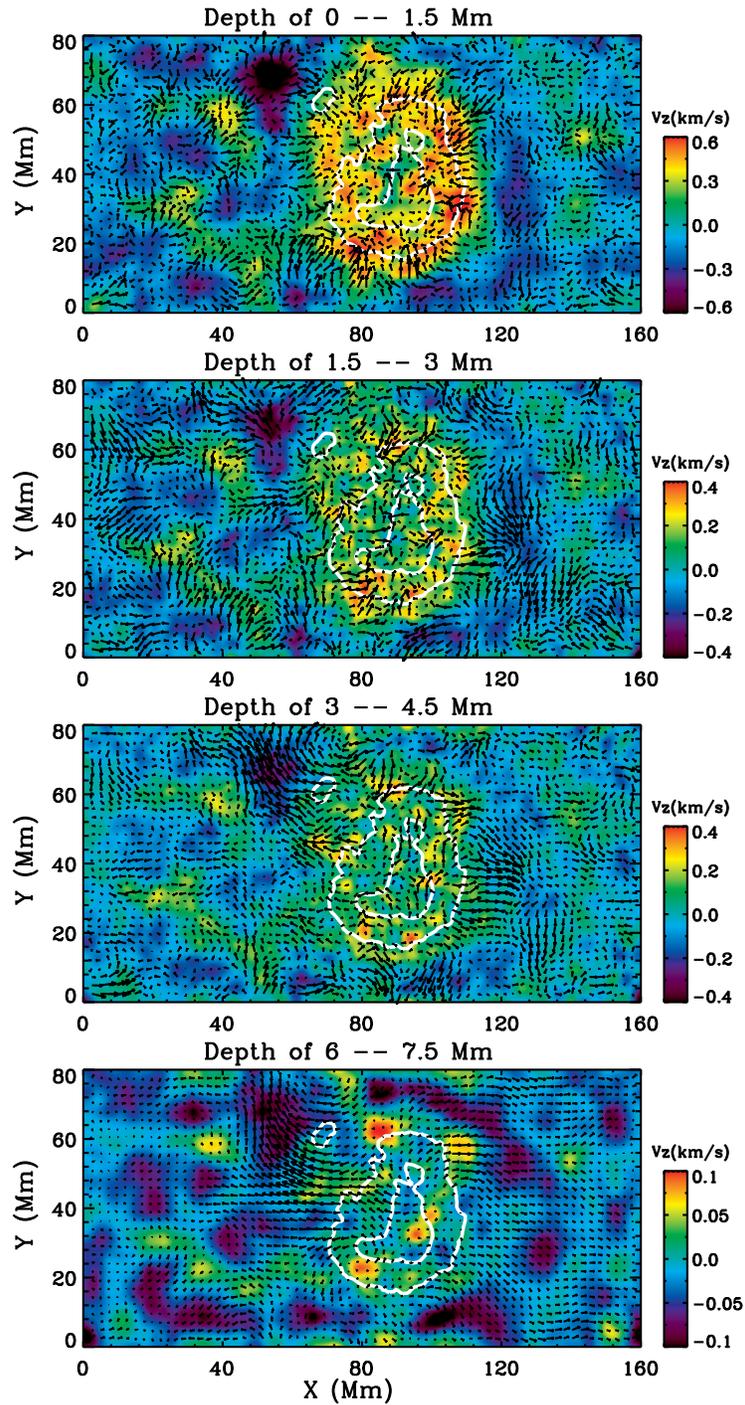}
\caption{Subsurface flow fields at different depths. Background images
indicate vertical flow fields, with positive representing downflows and
negative as upflows. Arrows indicate horizontal flows, with the longest
arrow representing a velocity of approximately 500 m/s in each panel.
White contours in each panel display the boundaries of sunspot umbra
and penumbra. }
\label{flows}
\end{figure}

A few selected three-dimensional velocity maps are displayed in 
Figure~\ref{flows} after a $4 \times 4$ rebinning in horizontal velocities 
in order to reveal more clearly the larger-scale patterns. It can be 
found that at the depth of 0 to 3 Mm, the plasma flows towards the 
boundary of sunspot penumbra from outside. Inside the sunspot 
penumbra, the horizontal flow fields are not clearly organized in
the rebinned images, but look well organized in the original high resolution 
flow maps (see \S3.3). Beneath 3 Mm, the horizontal flow field is largely 
divergent from the sunspot area, and extends quite far away. The vertical 
flows are directed mostly downward until at least 4.5 Mm in depth 
inside the sunspot area. Below 6 Mm or so, there seems to be a mixture of 
the downward and upward flows in this region.

\begin{figure}
\epsscale{0.8}
\plotone{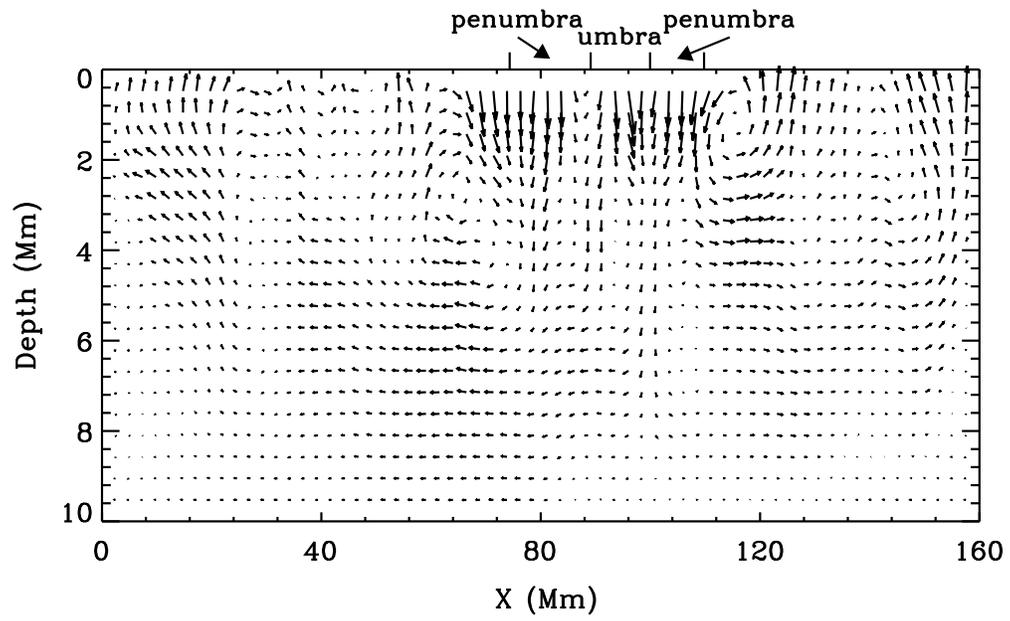}
\caption{A vertical view of flow field beneath the active region. The longest 
arrow is approximately 500 m/s. }
\label{vert_flows}
\end{figure}

A vertical view of the flow field (Figure~\ref{vert_flows}) after
linear interpolation shows nicely
the flow structure beneath the active region. Strong downdrafts are 
seen immediately below the sunspot's surface, and extends up to 6 Mm
into the deep. The downdraft has a maximum speed larger than 500 m/s, and 
is much stronger than the vertical flow speed expected in supergranular 
structures. A little beyond the sunspot's boundary, one can find 
both upward and inward flows. Clearly, large-scale mass circulations form 
outside the sunspot, bringing plasma down along the sunspot's boundary, 
and back to the photosphere within about twice of the sunspot's radius.  
It is remarkable that such an apparent mass circulation is obtained 
directly from the helioseismic inversions without applying any additional 
inversion constraints, such as forced mass conservation. The mass 
circulation pattern in the previous MDI results \citep{zha01} was 
not that clear.

Based on analysis of artificial data, \citet{zha03} demonstrated 
that a divergent (convergent) flow structure would result in a downward
(upward) vertical flow due to cross-talk effect, making the inversion
of small vertical speed difficult or even impossible by our inversion
technique. The same effect was also discussed by \citet{zha07}
to explain why they could not invert vertical flows from the numerical
simulations of the quiet Sun. However, the downward flows found in this
study is not due to the cross-talk effect, because the inflows around
active region found in this study would have caused upward speed due to
this effect. This means that in our results the inferred downward flows
are probably underestimated although it is not clear by how much. 
At the moment, we do not know how to offset this effect in our inversions.
This requires more work with sunspot numerical simulation data. However,
it is important to emphasize that this effect does not affect the 
qualitative picture of the flow pattern below the sunspot obtained in 
our inversions. 

\subsection{High Resolution Subsurface Flow Field}
Due to the very high spatial resolution of {\it Hinode} observation, 
the time-distance analysis can also give high resolution flow 
fields. One example, the horizontal flow field at $0 - 1.5$ Mm beneath 
the sunspot area, is shown in Figure~\ref{hr_flows}. From this flow map, 
one can easily identify small-scale divergent flow structures inside 
both the sunspot umbra and penumbra. The scale of these flow 
structures, $4 - 5$ Mm, is larger than the typical scale of solar 
granulations ($1 - 2$ Mm), but smaller than supergranulations ($20 
- 30$ Mm). The speed of these flows is about $300 - 500$ m/s. They 
probably represent convective cells in subsurface magnetized plasma
of sunspots. It is intriguing that small-scale vortex motions below 
sunspots were suggested by \citet{par92} as a key element of the cluster
sunspot model. 

However, one should also be cautious in interpreting these small-scale 
structures. The scale of these structures is similar to the wavelength 
of acoustic waves used for the inferences at this depth, approximately 
3 Mm. At this point, it is not clear how well the 
structures of this scale are resolved in our measurements. But the
evidence of small-scale structures below sunspots is quite interesting,
and certainly requires more observational and modeling studies. 

Another remarkable phenomenon in this high resolution flow map is that in
the light bridge area, located inside the upper part of the sunspot 
umbra (at $X = 28 - 32$ Mm and $Y = 33 - 37$ Mm), the flows are divergent 
from the bridge with a speed of $\sim 400$ m/s. This shows a strong 
coupling between the sunspot structure and subsurface dynamics. Thus,
the high resolution subsurface flow can be useful to monitor the dynamics 
and evolution of sunspots. 

\section{Analysis without Phase-Speed Filtering}
\subsection{Averaged Travel Times}
It has long been recognized from analysis of {\it SOHO}/MDI data that
for short distances, the time-distance cross-covariance functions from 
measurements may be corrupted by some signals that are not well understood 
\citep{duv97}. In Figure 7, we compare the time-distance diagrams calculated 
for the Hinode observing run and for high-resolution MDI data of 
the same duration and the same area size. 
As shown in the left panel of Figure~\ref{td_comp}, for the high
resolution MDI observations, when the distance is smaller than $\sim 8$ Mm, 
the acoustic wave propagation signal is essentially lost in some horizontal 
stripes, which block the measuring of acoustic travel times in this distance 
range. It is not yet well understood how these horizontal stripes are 
generated, but they are thought to be related to the instrument 
modulation transfer function (MTF) and instrumental distortions.
Phase-speed filtering was introduced 
to reduce the influence of these horizontal stripe signals at short distances
\citep{duv97}, but prior to the {\it Hinode} data there is yet no way to 
investigate the effectiveness and accuracy of this procedure. 

With the availability of {\it Hinode} high resolution helioseismology 
observations, this difficulty can be resolved or partly resolved. In 
the {\it Hinode} time-distance diagram shown in the right panel of 
Figure~\ref{td_comp}, the acoustic signals are very nicely visible in 
very short distances as small as $\sim2$ Mm, although 
some shorter and weaker horizontal stripes still exist. Judging from
this time-distance diagram averaged from the observations in the whole 
field of view as displayed in Figure~\ref{spot}, it should be possible 
to make the travel time measurements for travel distances shorter than 
4 Mm even without using the phase-speed filter.

Without using the phase-speed filter (but with $f$-modes and solar 
convection filtered out), we obtain three different time-distance diagrams
when the central points of the annulus are located inside a quiet 
solar region, sunspot penumbra, and sunspot umbra, respectively. These 
time-distance diagrams are obtained after averaging the outgoing and ingoing,
i.e., both positive and negative lags of, cross-covariance functions.
For a better comparison the quiet solar region is chosen a 
size similar to the sunspot umbra. The Gabor wavelet fitting 
\citep{kos96b} is used to infer the acoustic travel times for different 
distances and for all three diagrams, and the results are shown in 
Figure~\ref{tds}. The time-distance diagram looks quite noisy for the 
sunspot umbra, and less noisier for the sunspot penumbra. However, 
the acoustic wave propagation signals are strong and clear in both cases, and a 
robust fitting is not a problem for both diagrams.

It can be found that the measured acoustic travel times are longer
than the ray theory expectations at distance of about $6 - 7$ Mm.
This may be due to that the second-skip signals are not completely separated
from the first-skip signals, and the fitted values are thus slightly 
elevated. Since the travel time measurements inside the sunspot and 
in the quiet Sun should be similarly affected in the fitting, it is 
useful to use the quiet Sun measurements as a reference at this distance.
To better understand the travel time anomalies inside sunspots, we 
use the travel time measured in the quiet Sun as a reference, and 
study the relative travel times after the reference is subtracted
from the travel times measured in the sunspot umbra and penumbra (see 
Figure~\ref{td_diff}a and \ref{td_diff}b). It is clear that acoustic 
travel times measured inside the sunspot umbra and penumbra are quite
similar, and both exhibit a positive anomaly of $\sim40$ sec when 
the distance is smaller than approximately 12 Mm, and a negative 
anomaly of $\sim35$ sec when the distance is between approximately 14
and 21 Mm. 

For the correct interpretation of the travel time measurements, it is 
extremely interesting to compare these measurements with and without the 
phase-speed filter. For the measurements with the phase-speed 
filtering, we average the mean travel times, which are measured by use 
of 7 different annulus ranges using different phase-speed parameters 
as presented in \S3, inside the sunspot umbra and penumbra separately. 
Figures~\ref{td_diff}c and ~\ref{td_diff}d display such a comparison. 
It can be found that for both sunspot umbra and penumbra, the measurements 
with and without the phase-speed filtering are quite similar, i.e., 
the travel time anomalies are positive for travel distances shorter than 
$\sim13$ Mm, but become negative for larger distances. However, it appears 
that for the sunspot umbra, the measurements with the phase-speed filtering 
underestimate the values by a factor of $20 - 40\%$; and for
the sunspot penumbra, the values are also underestimated, but by a 
smaller factor. This underestimation is expected due to a combination
of the smaller oscillation amplitude in active regions and the use of 
phase-speed filtering \citep{raj06, par08}.

\subsection{Travel Time Map}
We have compared the averaged one-dimensional mean travel time 
curves with and without the phase-speed filtering, and it is interesting
to compare two-dimensional acoustic travel time maps that are obtained 
from two sets of different measurements. As it is well known that travel
time measurements without filtering are noisier, we have succeeded in making 
clean travel time maps for only three annuli among the seven that
are presented in \S3. Figure~\ref{map_comp} shows one such example 
of comparing mean travel time maps with and without the filtering. 
There are many similarities as well as discrepancies between these two
travel time maps. It is clear that at this measurement distance, for
both measurements, the negative travel time anomalies are found outside 
the sunspot area, and mixed positive and negative time anomalies exist
inside the sunspot. However, it is also clear the detailed structure of 
the map look different. Such comparisons are still at an early 
stage, and we believe a systematical study over such comparisons would 
lead us to a better understanding how phase-speed filtering, as well
as other filtering practices, effect measurements. Obviously, longer 
time series of high-resolution sunspot observations will help to 
improve the results. 

\section{Discussions and Summary} 
\subsection{Subsurface Structure and Flow Field} 
By a time-distance analysis of unprecedented high spatial resolution 
observations from the {\it Hinode}/SOT, we have investigated high resolution 
wave-speed structures and mass flows beneath active region AR10953. 
For the subsurface wave-speed structure, the inverted results are 
remarkably similar to previous results based on MDI Dopplergrams with 
different inversion technique as well as different sensitivity kernels. 
For subsurface flow fields, the general picture is similar to what has 
been obtained from MDI observations, i.e., converging 
downward flows near the surface below the sunspot area. Despite 
the similarities, the current picture also has some differences
compared with the earlier one. One difference is that the downward flows 
beneath the sunspot are more prominent in the flow fields inferred 
from this study. Another remarkable result from this study is the mass 
circulation outside the sunspot, which seems to keep mass conservative
and is much more clear than the previous MDI inversions. 
It is widely believed in theory and in numerical simulations that such 
downdraft and converging flows near the sunspot surface play important 
roles in keeping sunspots stable \citep{par79, hur00, hur08}. 

However, it is still not quite clear how the overall flow structures
around the sunspot's surface and interior look like. It is already
well known the existence of penumbral Evershed outflows, outgoing
moat flows beyond sunspot's penumbra, and inflows from the inner 
penumbra and umbra measured by tracking penumbra/umbra dots \citep[e.g.,][]
{sob09}. For the sunspot's interior, there were reports
of outflows from f-mode analysis \citep{giz00}, inflows at the
depth of 1.5 - 5 Mm (Zhao et al. 2001), and large scale inflows around
active regions up to 10 Mm in depth from ring-diagram analysis \citep{hab04}. 
Based on all these observations, \citet{giz03} proposed a 
schematic flow structure of the sunspot, and \citet{hin09} recently 
proposed a similar one. Both flow structures
show outflows near the surface, and inflows below that, with transition
depth of these opposite flows undetermined. Here, based on all previous
results, as well as on the recent numerical simulation of Evershed flows
\citep{kit09}, we also propose a schematic flow structure 
of a sunspot (Figure~\ref{flow_profile}), slightly different from the
plots of two previous studies, but essentially similar. However,
it is also acknowledged that this flow structure is not consistent 
with what was recently found by \citet{giz09} and recent numerical
simulations by \citet{rem09}. Thus, further observational and
theoretical studies are required to determine the subsurface dynamics
of sunspots. 

\begin{figure}
\epsscale{0.9}
\plotone{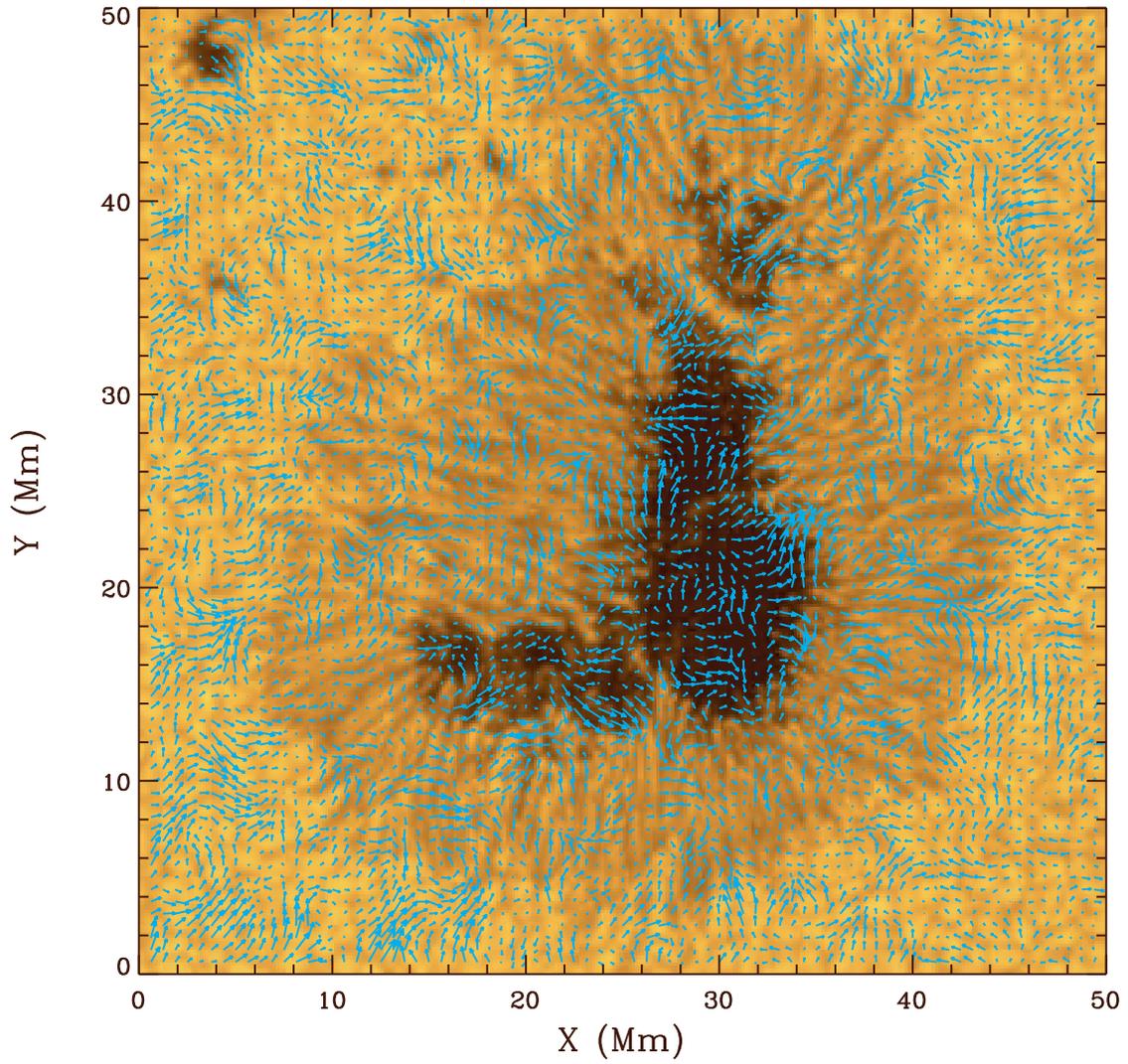}
\caption{High resolution horizontal flow field 0 - 1.5 Mm beneath the sunspot
region. The background image is this sunspot observed at the photospheric
level. The longest arrow indicates a speed of 800 m/s. }
\label{hr_flows}
\end{figure}

\begin{figure}
\epsscale{0.9}
\plotone{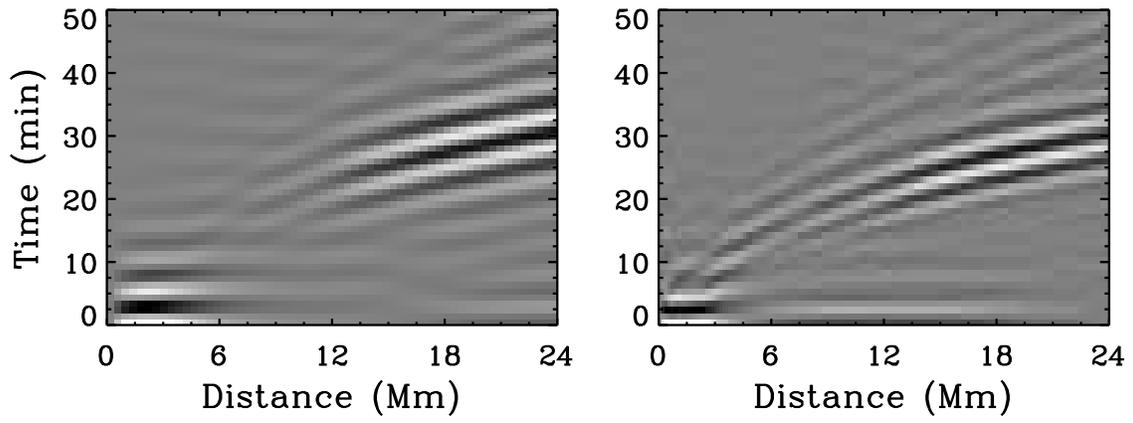}
\caption{Time-distance diagrams obtained from MDI high-resolution Dopplergram
({\it left}) and {\it Hinode} ({\it right}) \ion{Ca}{2}~H observations. }
\label{td_comp}
\end{figure}

\begin{figure}
\epsscale{1.0}
\plotone{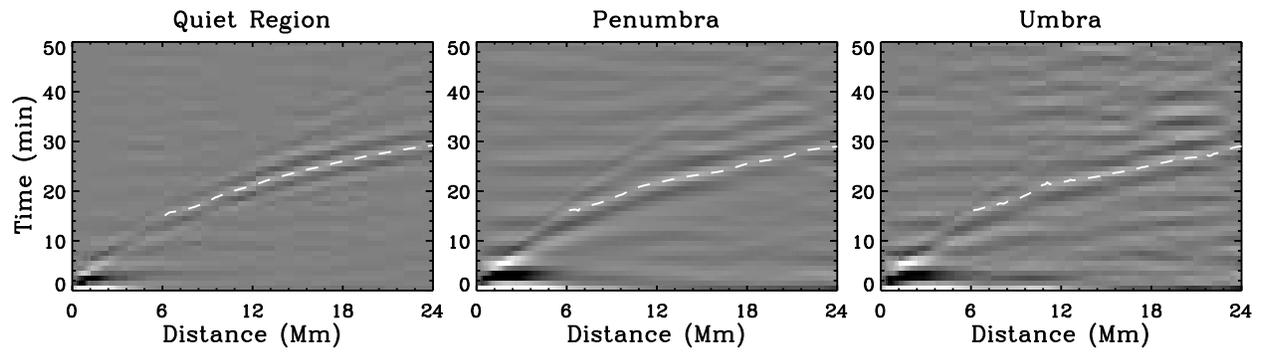}
\caption{Time-distance diagrams obtained when the central points are located
inside a quiet region, sunspot penumbra, and sunspot umbra, respectively. 
The dashed line in each panel is the travel times fitted from the 
corresponding time-distance diagram. } 
\label{tds}
\end{figure}

\begin{figure}
\epsscale{0.9}
\plotone{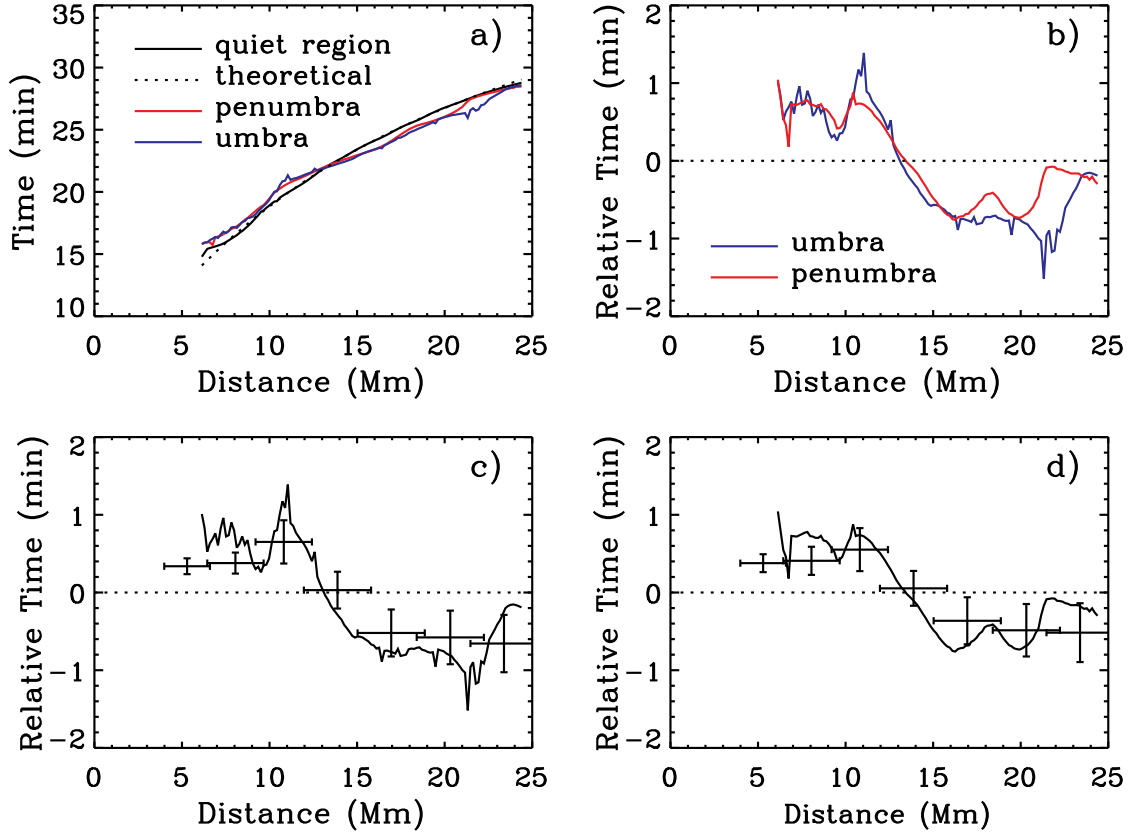}
\caption{a) Comparison of travel times obtained in solar quiet region, sunspot
penumbra, and sunspot umbra. Solid lines are the same as dashed lines in 
Figure~\ref{tds}, and the dotted line is a theoretical expectation of travel
times from ray theory. b) Travel time differences relative to the quiet region
for sunspot penumbra and umbra. c) Comparison of travel time differences
with and without using phase-speed filtering. Solid line is the same as 
the umbra line in b), and the points with error bars are mean travel times 
averaged from the travel time maps obtained with the use of phase-speed 
filtering (i.e., measurements used in \S3), but only points inside sunspot umbra
are used for average. Horizontal error bars indicate the annulus radius 
range, and the vertical error bars indicate the standard errors. d)
Same as panel c), but solid lines and points with error bars are both
obtained inside sunspot penumbra. }
\label{td_diff}
\end{figure}

\begin{figure}
\epsscale{0.9}
\plotone{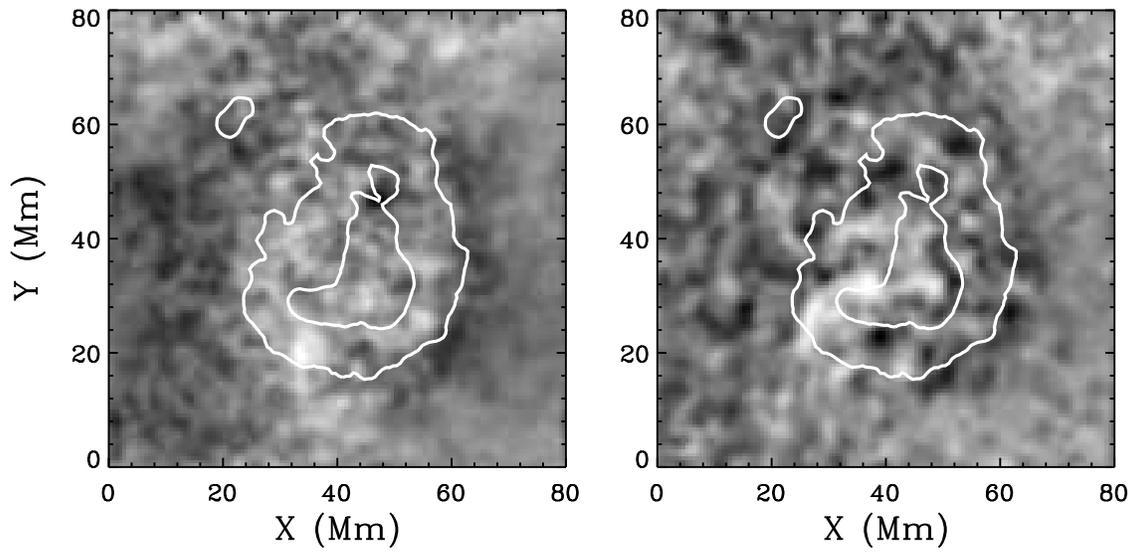}
\caption{Comparison of mean travel time maps obtained with ({\it left}) and 
without ({\it right}) the phase-speed filtering. The annulus radius range
is $12.0 - 15.9$ Mm, same for both measurements. White contours indicate
the boundaries of sunspot umbra and penumbra. }
\label{map_comp}
\end{figure}

The high resolution subsurface wave-speed maps (Figure~\ref{cs_horiz})
and flow fields (Figure~\ref{hr_flows}), present us an unprecedented 
opportunity to see the sunspot's subsurface structures and dynamics with 
many details. These results reveal the complexity of the subsurface 
perturbations and dynamics and their relationship with the sunspot
structure. The subsurface image at the depth of $0 - 1.5$ Mm displays a larger 
wave-speed perturbation, presumably hotter in temperature, near the light 
bridge area than other areas inside the sunspot, where the wave-speed
perturbations are largely negative. This is in agreement with the 
surface observations that light bridges are formed by upwelling of hot 
plasma \citep{kat07}. We also see clear divergent flows from the light 
bridge in high resolution subsurface flow map at the same depth 
(Figure~\ref{hr_flows}). This is consistent with the plasma upwelling 
that occurs in this area, but not in good agreement with the 
proper motion tracking results of this region \citep{lou08}. 

It is also interesting to see quite a few divergent flow cells inside both 
the sunspot umbra and penumbra. The cells are bigger than granules
and smaller than supergranules, and indicate the existence of a 
mesoscale convection inside the sunspot. The discovery of these motions 
is intriguing because they may be a signature of the downdraft vortex 
rings around magnetic flux bundles, as suggested by \citet{par92}. 
The Parker's conjecture was that the observed clustering of magnetic 
flux bundles in sunspot is at least in part a consequence of 
hydrodynamic attraction of the downdraft vortex rings. Thus, it is 
very important to continue the investigations of the subsurface 
dynamics of sunspots, both observationally by high resolution 
helioseismology and theoretically by realistic MHD simulations.

\subsection{Analysis without Phase-Speed Filtering} 
As introduced in \S1, the phase-speed filtering may introduce 
some errors in active regions due to the smaller oscillation amplitude
in these areas \citep{raj06}, and using ridge filtering may give 
different results as using phase-speed filtering in active regions
\citep{bra08}. All of these prompt us a reexamination 
of the use of phase-speed filtering and an evaluation of how phase-speed
filtering alter measured acoustic travel times, in particular, 
inside active regions. 

The {\it Hinode} high resolution observation has helped us to overcome 
an obstacle that acoustic travel times could hardly be measured in
short distances inside active regions if phase-speed filtering was not 
applied. Our measured distance-dependent acoustic travel time differences 
relative to the solar quiet region obtained without using the phase-speed 
filtering, as shown in Figure~\ref{td_diff}, have evidently demonstrated 
that for short distances, the acoustic travel time is 
longer inside both sunspot umbra and penumbra than the quiet 
region, and for distances longer than $\sim 13$ Mm, the mean travel time
is shorter. This is generally in agreement with the measurements using 
phase-speed filters, except that the ones with filters would underestimate
the values by $20-40\%$ in sunspot umbra. It is believed that the 
underestimation is caused by a combination of oscillation amplitude 
suppression inside active regions and the use of phase-speed filtering.
Therefore, we believe that the phase-speed filtering does not change 
qualitatively the measured travel times, therefore the inferred interior 
properties of active regions, but would change the inferred values 
quantitatively. 

The {\it Hinode} data also enable us to construct acoustic travel time
maps without using filters, and this gives us an opportunity to assess
the correctness and accuracy of various filters. Further investigations
are particularly important in this regard.

\subsection{Summary}
We summarize our results as follows:
\begin{enumerate}
\item By analyzing high resolution {\it Hinode} \ion{Ca}{2}~H observations
using time-distance helioseismology, we have derived wave-speed structures 
beneath solar active regions. The inferred structures are 
remarkably similar to results obtained by various authors 
analyzing MDI Dopplergram observations,
although the instrument, the spectrum line used to observe, the type
of observation (intensity and Doppler velocity), the inversion technique, 
and the sensitivity kernels are mostly different. 
\item The subsurface flow field structure is generally in agreement with
previous MDI result, but it is clear that the downward flows near
the sunspot surface is more prominent, and the mass circulation around
the sunspot is more clear and seems self-consistent, although no
mass conservation constraint is applied in the inversion.
\item Our analysis without using phase-speed filtering has convincingly
demonstrated that the phase-speed filtering does not change 
travel time measurements qualitatively, but may underestimate travel
time anomalies inside active regions.
\item High spatial resolution subsurface flow fields reveal quite a few
organized flow structures inside sunspot umbra and penumbra, which are
perhaps corresponding to some convection structures or downdraft 
vortex rings around magnetic flux bundles. In the light bridge 
area, subsurface hotter temperature is found, and
plasma flows divergent from the light bridge is also seen.
These initial results provide a basis for further development of 
high-resolution acoustic tomography of sunspots, and comparison
with numerical simulations. 
\end{enumerate}

\begin{figure}
\epsscale{0.7}
\plotone{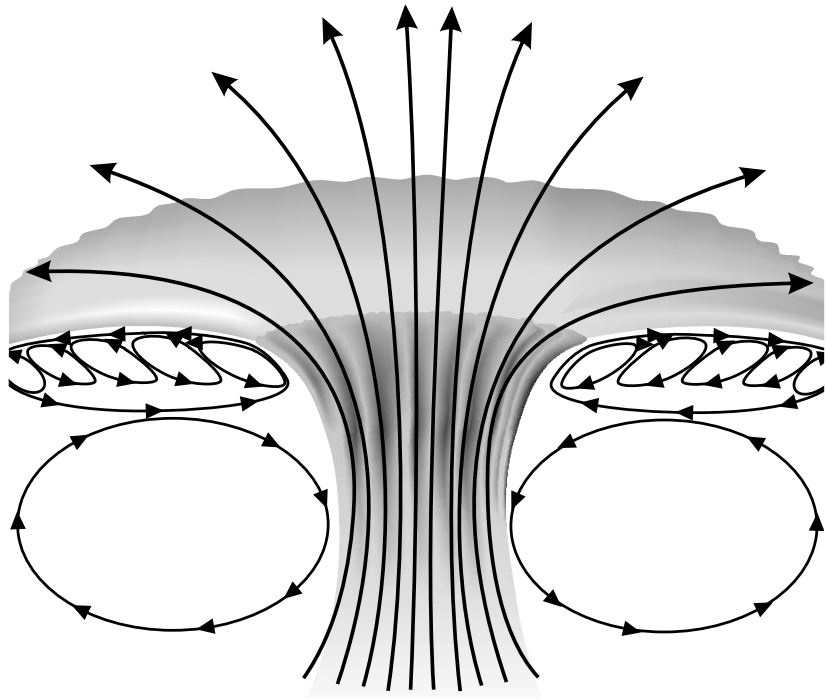}
\caption{A schematic plot of sunspot's flow structure near the surface and 
in the interior. }
\label{flow_profile}
\end{figure}

\acknowledgments
We thank Dr.~I.~Kitiashvili for making Fig.~\ref{flow_profile}
for us. We also appreciate an anonymous referee for raising a few
important questions that helped to improve the quality of this
paper. {\it Hinode} is a Japanese mission developed and launched by
ISAS/JAXA, collaborating with NAOJ as a domestic partner,
NASA and STFC (UK) as international partners. Scientific
operation of the {\it Hinode} mission is conducted by the {\it Hinode}
science team organized at ISAS/JAXA. This team mainly consists
of scientists from institutes in the partner countries.
Support for the post-launch operation is provided by JAXA and
NAOJ (Japan), STFC (U.K.), NASA, ESA, and NSC (Norway).
This work was (partly) carried out at the NAOJ {\it Hinode} Science
Center, which is supported by the Grant-in-Aid for Creative
Scientific Research ``The Basic Study of Space Weather
Prediction'' from MEXT, Japan (Head Investigator: K.~Shibata),
generous donations from Sun Microsystems, and NAOJ internal
funding.

\clearpage

\clearpage

\clearpage

\clearpage

\clearpage

\clearpage

\clearpage

\clearpage

\clearpage

\clearpage

\clearpage

\clearpage

\clearpage

\end{document}